# Modeling Majorness as a Perceptual Property in Music from Listener Ratings

Anna Aljanaki,[1] Gerhard Widmer[2]

*Institute of computational perception, Johannes Kepler University, Austria*

[1]aljanaki@gmail.com, [2]gerhard.widmer@jku.at

## Abstract

For the tasks of automatic music emotion recognition, genre recognition, music recommendation it is helpful to be able to extract mode from any section of a musical piece as a perceived amount of major or minor mode (majorness) inside that section, perceived as a whole (one or several melodies and any harmony present). In this paper we take a data-driven approach (modeling directly from data without giving an explicit definition or explicitly programming an algorithm) towards modeling this property. We collect annotations from musicians and show that majorness can be understood by musicians in an intuitive way. We model this property from the data using deep learning.

## Introduction

With Western popular tonal music, the term "mode" is often used dichotomously to refer only to major (Ionian) or minor (Aeolian) mode (omitting the rest of the modes, such as harmonic minor, blues scales, etc.). For a certain combination of harmony and melody, labeling an excerpt of music as "major" or "minor" can be subjective, ambiguous, or even impossible (especially when modulations and/or key signature changes are present inside the segment). The tonal hierarchy also needs time establish itself (Parncutt, 1989) and the perception of the tonal centroid and mode may change while this is happening. In music information retrieval context, the category of mode can sometimes be treated probabilistically (e.g., as a probability of an excerpt being in major mode, as predicted by an algorithm), resulting in a continuous property (Saari, 2011) and (Friberg, 2011).

In this paper we will call this property majorness, following (Parncutt, 1989) and MIRToolbox (Lartillot, 2008). There exist music analysis tools that permit to extract it, such as MIRToolbox and QM Vamp Plugin (Noland, 2007). The algorithms implemented in these tools rely on a pitch class profile based estimation of key, and produce a result that is not exactly similar to a perceptual estimation of majorness, as annotated by musicians (Friberg, 2011).

The concept of majorness lacks a clear musicological definition, and therefore it is difficult to design an algorithm to extract it in the same straightforward way as an algorithm for onset detection can be designed. However, given enough training data, it is possible to learn the property directly from the data. In this paper we investigate this approach.

## Methods

In this section, we will first describe our data collection approach (a hybrid of pairwise comparison and absolute ranking), and then describe the deep learning method that was used to create a model of majorness.

## Data collection

It is very difficult for an annotator to rate a vaguely defined and subjective concept such as majorness on an absolute scale (Madsen, 2013). Comparing two examples given a certain criterion is an easier task. However, pairwise comparisons require factorially (in relation to the number of examples) more ratings, as compared to linear number of absolute ratings, even when using only a part of the full comparison matrix (Madsen, 2013). In order to learn majorness from data, we need to annotate at least several thousand song excerpts, this amount of songs would require millions of pairwise comparisons, which is prohibitively expensive.

We decided to combine the two approaches and first create a scale using pairwise comparisons, and then collect absolute ratings on that scale.

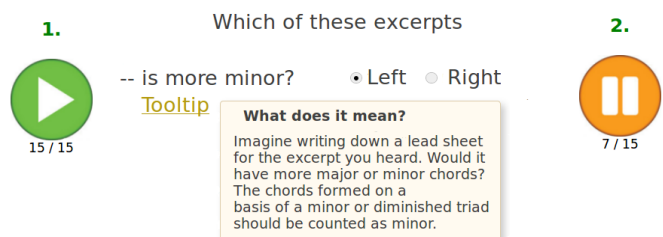

**Figure 1.** Pairwise comparisons interface with a hovered tooltip (translated from a Russian interface on toloka.yandex.ru).

### Pairwise comparisons

On a crowd-sourcing platform we hire 80 musicians (5 per pair) to compare pairwise 100 musical excerpts of 15 seconds on their majorness. We get the music from creative-commons licensed websites and chose 100 songs from different genres and with different valence/arousal values. Figure 1 shows the interface that was shown to the annotators on the Toloka crowd-sourcing platform (toloka.yandex.ru). From the pairwise comparisons, we obtain a ranking of pieces from the most certainly minor ones, through the ambiguous ones, to the most certainly major ones.

### Absolute rankings

From the ranking we sample 10 excerpts as examples and collect ratings of perceived majorness for 5000 excerpts, also belonging to various music genres (rock, pop, classical, jazz, blues, etc.). Figure 2 shows the interface that was used to collect these ratings. An annotator compares a piece to an example, and if a piece is more minor than an example, listens to the next example to the right, until the current piece can be placed between two examples (it's more minor than example to the left, and less minor than example to the right). In this way, absolute ratings from 1 to 10 are obtained.

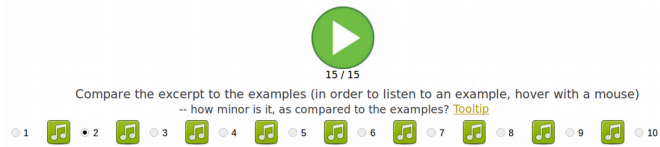
**Figure 2. An interface for absolute scale ratings (translated).**

### Deep learning model

From every musical excerpt in the dataset we extract a mel-spectrogram with 299 mel-filters with a half-overlapping Hanning window of 2048 (44.1k sampling rate). We train a fully convolutional neural network (Inception architecture) with a mean squared error loss (regression task) on the averaged absolute ratings. More details about the model can be found in (Aljanaki, 2018).

## Results

### Data

The consistency of the annotations without any unreliable rater filtering is 0.69 Cronbach's alpha (0.33 Krippendorff's alpha). These valuse indicate low consistency. To improve the annotations, some of the annotators were removed based on their disagreement with the rest. Figure 3 shows a histogram of the resulting annotations. The data is normally distributed, showing that the annotators avoided the extremes (completely major and completely minor).

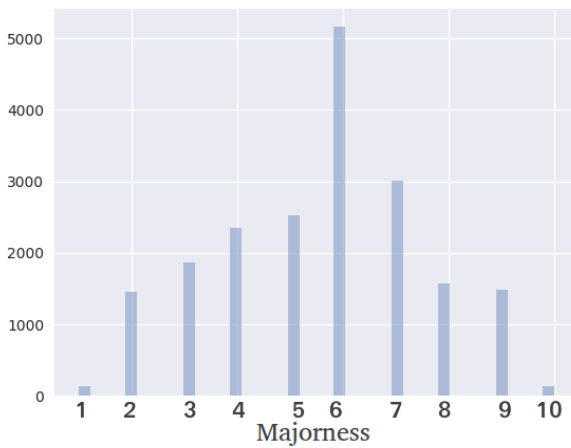
**Figure 3. Distribution of the annotations for majorness.**

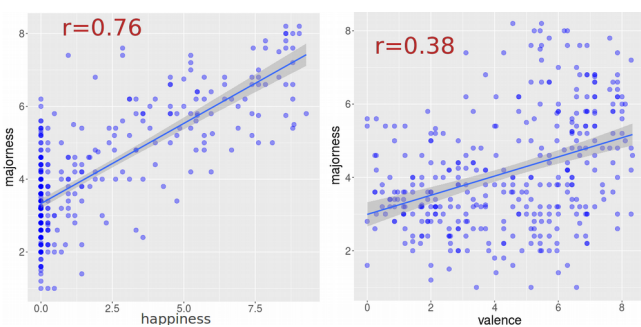
**Figure 4. Correlation of majorness with valence and happiness.**

It has been shown that majorness is a useful feature for predicting emotion in music (Gabrielsson, 2001). We included the songs from a soundtracks dataset annotated with emotion, both dimensional (valence and arousal) and categorical (five basic emotions, which are used both as categories and as dimensions) (Eerola, 2011). Figure 4 shows the correlation of majorness as annotated in our dataset with valence and happiness (used as a dimension) on the 360 songs from the soundtracks dataset. There is a strong correlation with happiness, which is expected from majorness. The correlation with valence is less strong.

### Predicting mode on WTC

An Inception model trained on this data, as explained before, could predict majorness on the test set with a Pearson's correlation of 0.48. In case of a neural network, it is difficult to understand, what exactly the model has learned. In order to better understand it, we used a collection of pieces in different tonalities – the Well-Tempered Clavier (WTC) by J.S. Bach (both books). The WTC contains 96 preludes and fugues, 48 major and 48 minor ones. We used recordings of Glenn Gould's performances and extracted the mel-spectrograms from the first 12 seconds of each prelude or fugue.

With a continuous majorness feature predicted by the model as an independent variable, we train a logistic regression to predict binary major and minor mode. This model has 70% accuracy when 10-fold cross-validated (random baseline is 50%). Clearly, what a model has learned is related to mode, but not exactly mode.

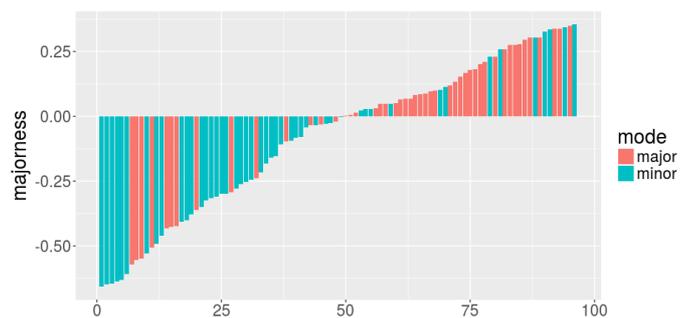
**Figure 5. Majorness vs major and minor mode in WTC.**

Figure 5 shows how the algorithm performed on minor and major pieces from WTC. The order from left to right is not corresponding to the order of the pieces in WTC, but to the majorness predicted by the neural network. By examining the mistakes (red rows below the x axis, and blue rows above the x axis), we can see that slower and more pensive performances of major pieces were classified as minor (prelude from BWV852, fugue from WV892), and faster, more energetically performed minor pieces were classified as major (prelude from BWV875, prelude from BWV871).

## Conclusion

Even without being given a formal definition of majorness, musicians could somewhat intuitively understand this property, and agree on it when annotating music. The property that they annotated turned out to be very close to valence (happiness) dimension of musical emotion. Using this data, we trained a neural network to predict majorness from musical audio. It appears that except for mode itself, a neural network also takes into account other (perhaps, more performative) aspects.

**Acknowledgements.**

This work was supported by the European Research Council (ERC) under the EUs Horizon 2020 Framework Programme (ERC Grant Agreement number 670035, project "Con Espressione"). This work was also supported by FCS grant.